\documentclass{PoS}

\usepackage{amsmath}
\usepackage{epsfig}
\usepackage{subfigure}

\PoS{PoS(LAT2005)252}

\title{Evaporation/Condensation of Ising Droplets\thanks{Work
    supported by the Graduiertenkolleg ``Quanten\-feld\-theorie:
    Mathematische Struktur und Anwendungen in der Elementarteilchen-
    und Festk\"orperphysik'' and by the Deutsche Forschungsgemeinschaft
    (DFG) under grant No. JA483/22-1.} }

\ShortTitle{Evaporation/Condensation of Ising Droplets}

\author{\speaker{Andreas Nu\ss baumer}, Elmar Bittner and Wolfhard Janke\\
  Institut f\"ur Theoretische Physik,\\
  Universit\"at Leipzig,\\
  Augustusplatz 10/11,\\
  D-04109 Leipzig, Germany\\
  E-mail: \email{andreas.nussbaumer@itp.uni-leipzig.de}}

  \abstract{ Recently Biskup {\em et~al.\/} [{\em Europhys.\ Lett.}
    {\bf 60} (2002) 21] studied the behaviour of $d$-dimensional
    finite-volume liquid-vapour systems at a fixed excess $\delta N$
    of particles above the ambient gas density. They identify a
    dimensionless parameter $\Delta (\delta N)$ and a universal
    constant $\Delta_\mathrm{c}(d)$ and show that for $\Delta <
    \Delta_c$ a droplet of the dense phase occurs while for $\Delta >
    \Delta_c$ the excess is absorbed in the background.  The fraction
    $\lambda_\Delta$ of excess particles forming the droplet is given
    explicitly. Furthermore, they state, that the same is true for
    solid-gas systems.

    To verify these results, we have simulated the spin-$1/2$ Ising
    model on a square lattice at constant magnetisation equivalent to
    a fixed particle excess in the lattice-gas picture. We measured
    the largest minority droplet, corresponding to the solid phase, at
    various system sizes ($L=40, \dots, 640$). Using analytic values for
    the spontaneous magnetisation $m_0$, the susceptibility $\chi$ and
    interfacial free energy $\tau_\mathrm{W}$ for the infinite system,
    we were able to determine $\lambda_\Delta$ in very good agreement
    with the theoretical prediction.  
  }
  
  \FullConference{XXIIIrd International Symposium on Lattice Field Theory\\
    25-30 July 2005\\
    Trinity College, Dublin, Ireland}

\begin{document}

\section{Introduction}
One of the longstanding problems in statistical mechanics concerns the
formation and dissolution of equilibrium droplets at a first-order
phase transition. Interesting quantities in this context are the size
and the free energy of a ``critical droplet'' that needs to be formed
before the decay of the metastable state via homogeneous nucleation
can start.  In the analysis done so far it was implicitly assumed that
the size of the droplet is of the order of the system, which is not
the case when the droplet forms first \cite{dobrushin.shlosman,binder.kalos}.
In this work, the region of system parameters that lead
to the formation/dissolution of a droplet is examined by means of
Monte Carlo simulations. We follow closely the theoretical ideas of
Biskup {\em et~al.\/} \cite{kotecky}; still, there are also different
descriptions of the same physical phenomena \cite{binder.kalos, furukawa.binder,
neuhaus.hager,binder}.

In the remaining sections, after a brief introduction to the
classical droplet theory, first the new results of 
Biskup {\em et~al.\/} \cite{kotecky} are summarized. Then our Monte Carlo 
measurements supporting the theoretical results are presented, and finally 
some preliminary conclusions are drawn.

\section{Theory}
In the following the term ``droplet'' will be used in the sense of
particles (or spins) that are grouped together in a purely geometric
way.  One of the basic papers in this context is due to M.\ E.\ Fisher
\cite{fisher} where he discusses "... a gas of particles interacting with
repulsive cores and short-range attractive forces ...". There, he
mentions that "... the typical configuration at low densities and
temperatures will consist of essentially isolated clusters of one, two,
three or more particles" while "A sufficiently large cluster is just a
small droplet of the liquid ..." and "Condensation in this picture
corresponds to the growth of a macroscopic droplet of the liquid."
It should be emphasised that the {\em geometric\/} definition of a 
cluster or droplet must not be confused
with {\em stochastic\/} ``Fortuin-Kasteleyn clusters'' or anything 
the like where it is not possible to identify a cluster by just ``looking'' 
at a spin configuration.

All arguments in this section are based on the work of Fisher
\cite{fisher} and especially Biskup {\em et~al.\/} \cite{kotecky}. Due
to the fact that the Monte Carlo simulations were done using the Ising
model, the theory is presented in terms of a lattice gas which mainly
is a change of notation but does not alter the theory. This is also
supported by Biskup {\em et~al.\/}'s presentation where the general 
results are additionally given in the special context of the 
two-dimensional Ising model with Hamiltonian
%
%
\begin{equation}
  {\cal H} = -J \sum_{\left< i,j\right>} \sigma_i \sigma_j - h \sum_i
  \sigma_i \;, 
\end{equation}
where $\sigma_i=\pm 1$ and $\left<i,j \right>$ denotes a
nearest-neighbour pair.  Having no external field ($h=0$)
the second term vanishes. When an up-spin ($\sigma_i=1$) is treated as a
particle and a down-spin ($\sigma_i=-1$) is treated as a vacancy the
system can be interpreted as a lattice gas of atoms. 

Considering such a system, classical droplet theory assumes that there
are two contributions to the free energy. First, there can be local
fluctuations and the probability to find a difference in the
magnetisation (excess in the magnetisation compared to $M_0=m_0 V$
with $V = L^2$) of
$\delta M = M-M_0$ can be expressed in terms of a Gaussian
distribution as
\begin{equation}
  \label{eqn:fluctuations}
  \exp\left[- \frac{(\delta M)^2}{2 V \chi}\right]
  = \exp\left[- \frac{(2 m_0 v_L)^2}{2 V \chi} \right] .\;
\end{equation}
Here, $m_0 = m_0(\beta)$ is the spontaneous magnetisation and $\chi =
\chi(\beta)$ the susceptibility, both quantities at an inverse
temperature $\beta$ and in the thermodynamic limit.
\begin{figure}
  \begin{center}
    \includegraphics[clip,scale=0.85]{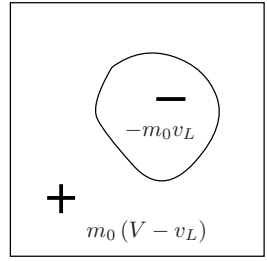}
  \end{center}
  \vspace*{-2mm}
  \caption{Ising system of size $V$ with a minority droplet of volume
    $v_L$ of negative spins surrounded by positive spins with an
    volume $(V-v_L)$. Here we have assumed for simplicity that the total excess in magnetisation
 	 is concentrated in the droplet, i.e. $v_d=v_L$.}
  \vspace*{-1mm}
  \label{fig:ising}
\end{figure}
Assuming that the magnetisation in the background and inside a droplet is
$m_0$, the total magnetisation can be written as $M=-m_0 v_L + m_0
(V-v_L)$, where $v_L$ is the volume of the droplet and $V$ the volume
of the system (see Fig.~\ref{fig:ising}). Rearranging this expression
for the difference in the magnetisation yields 
\begin{equation}
  \label{eqn:deltaM.vL}
  \delta M = M-V m_0 = -2 v_L m_0
\end{equation}
which was used in Eq.~(\ref{eqn:fluctuations}).

The second contribution to the free energy stems from an interface of a
droplet of volume $v_L$. The cost to form it is given in two dimensions 
by \cite{shlosman}
\begin{equation}
  \label{eqn:droplet}
  \exp\left[ -\tau_\mathrm{W} \sqrt{v_L} \, \right] \;,
\end{equation}
where $\tau_\mathrm{W}=\tau_\mathrm{W}(\beta)$ is the interfacial free
energy per unit volume of an ideal shaped droplet which is also known
as the free energy of a droplet of Wulff shape. 

\begin{figure}
  \subfigure[Evaporated system where a large number of very
  small clusters exist (1 to 3 spins)]{
    \label{fig:ce.a}
    \hspace{2cm}
      \includegraphics[clip,scale=1.5]{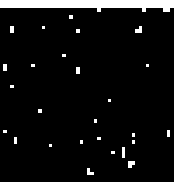}
    \hspace{2cm}
    }   
  \subfigure[Condensed system with a single large cluster that
  has absorbed nearly all small clusters]{
    \label{fig:ce.b}
    \hspace{2cm}
    \includegraphics[clip,scale=1.5]{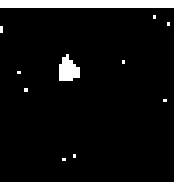}
    \hspace{2cm}
  }   
  \vspace*{-1mm}
  \caption{Snapshots of a two-dimensional Ising lattice; white squares
    correspond to up-spins (atoms), while black squares correspond to
    down-spins (vacancies).}
  \vspace*{-1mm}
  \label{fig.ce}
\end{figure}

Comparing the exponents of Eq.~(\ref{eqn:fluctuations}) to
(\ref{eqn:droplet}) gives:
\begin{equation}
  \label{eqn:delta}
  \Delta = \frac{(2 m_0 v_L)^2/(2 V \chi)}{ \tau_\mathrm{W} \sqrt{v_L}}
  = 2 \frac{m_0^2}{\chi \tau_\mathrm{W}} \frac{v_L^{3/2}}{V} \;.
\end{equation}
With $\Delta \stackrel{!}{=} 1$ and Eq.~(\ref{eqn:deltaM.vL}), the
difference in the magnetisation is 
\begin{equation}
  \delta M =  \theta V^{2/3} 
  \hspace{1cm}
  \text{with}
  \hspace{1cm}
  \theta = - \left(\frac{2 \chi \tau_\mathrm{W}}{\sqrt{2
        m_0}}\right)^{2/3} \;.
\end{equation}
This means, when $\delta M \gg \theta V^{2/3}$ the droplet mechanism
dominates, while the fluctuation mechanism dominates for $\delta M
\ll \theta V^{2/3}$. Biskup {\it et al.\/} \cite{kotecky} studied the
crossover region $\delta M \approx \theta V^{2/3}$. By isoperimetric
reasoning, they showed that in this range no droplets of intermediate
size exist. There is at most a single large droplet of size
$v_\mathrm{d} < v_L $ with costs
giving in Eq.~(\ref{eqn:droplet}) that absorbs $\delta M_\mathrm{d}$ of the excess
of the 
magnetisation $\delta M$ while the rest goes into the fluctuations of
the background. This justifies the following Ansatz for the
probability that the droplet contains the fraction $\lambda = \delta
M_\mathrm{d} / \delta M$ of the excess in magnetisation:
\begin{equation}
  \label{eqn:minimisation}
  \exp\!\left[ -\tau_\mathrm{W} \sqrt{v_\mathrm{d}}-\frac{(\delta M - \delta
  M_\mathrm{d})^2}{2 V \chi} \, \!\right]  
  = \exp\!\left[-\tau_\mathrm{W} \sqrt{\frac{-\delta M}{2 m_0}}
    \Phi_\Delta(\lambda) \!\right] 
  \, \text{,} \quad
  \Phi_\Delta(\lambda) = \left[\sqrt{\lambda} + \Delta (1-\lambda)^2
  \right] \, , 
\end{equation}
where $\Delta$ is defined in Eq.~(\ref{eqn:delta}). Since
$\tau_\mathrm{W} \sqrt{-\delta M /2 m_0}$ and $\Delta$ are
constants, the fraction of excess that is most probable is obtained by
minimising $\Phi_\Delta(\lambda)$. In two dimensions, the minimisation 
leads to
$\Delta_\mathrm{c}= (1/2)(3/2)^{3/2} \approx 0.92$. For values $\Delta
< \Delta_\mathrm{c}$ the global minimum of $\Phi_\Delta(\lambda)$ is
reached for $\lambda=0$, while for $\Delta > \Delta_\mathrm{c}$ it is
located at a nontrivial value $\lambda_\Delta > 0$. At the
transition point $\Delta=\Delta_\mathrm{c}$, the value is
$\lambda_\mathrm{c}=2/3$. The solid line in Fig.~\ref{fig:main.result} below
shows the graph of $\lambda_\Delta$. Its interpretation is as
follows: for $\Delta<\Delta_\mathrm{c}$ all of the excess is absorbed
in the background fluctuations, then, at the transition point
$\Delta=\Delta_\mathrm{c}$ a value of $2/3$ of the excess forms a
droplet while the rest of the excess remains as background
fluctuations. For values $\Delta > \Delta_\mathrm{c}$ the droplet
grows and absorbs most of the background fluctuations. For an illustration 
with actual simulation data, see Fig.~\ref{fig.ce}.

\section{Simulations}
The main goal of our work was to check the theoretical results
presented in the last section. In order to do so the fraction of the
excess of magnetisation in the largest droplet was measured.
To this end a Monte Carlo simulation was set up with a fixed excess of
magnetisation. To keep the magnetisation during the simulation
constant, a Metropolis update with Kawasaky dynamics was chosen.
After every sweep a cluster decomposition was performed (using the
Hoshen-Kopelman algorithm) and the volume of the largest cluster was
measured. It is important to note that in the present context
the volume of the cluster includes overturned spins within the
cluster (which was implemented with a so-called ``flood-fill'' routine).

The simulations were performed at 38 different magnetisations chosen
to have enough data points in the vicinity of the transition. The
temperature was set to $T=1.5$ and altogether five different square lattice
sizes were taken into account ($L=40, 80, \dots, 640)$. Every
simulation took $20\, 000$ sweep for thermalisation and $200\, 000$
sweeps for measurements. To obtain the error bars, 10 independent
simulations were run for each data point.

In order to get the correct scaling for the abscissa, $\Delta(M, M_0,
\chi, \tau_\mathrm{W})$ has to be calculated. For the spontaneous
magnetisation there exists Onsager's famous analytic solution and for
the free energy of the Wulff droplet, Rottman and Wortis \cite{wortis}
were able to derive an analytic expression, while for the
susceptibility only series expansions are known; see, e.g., Enting, Guttmann
and Jensen \cite{enting} (but up to order 323!).

\section{Results}

\begin{figure}
  \begin{center}
    \includegraphics[clip,clip,scale=0.85]{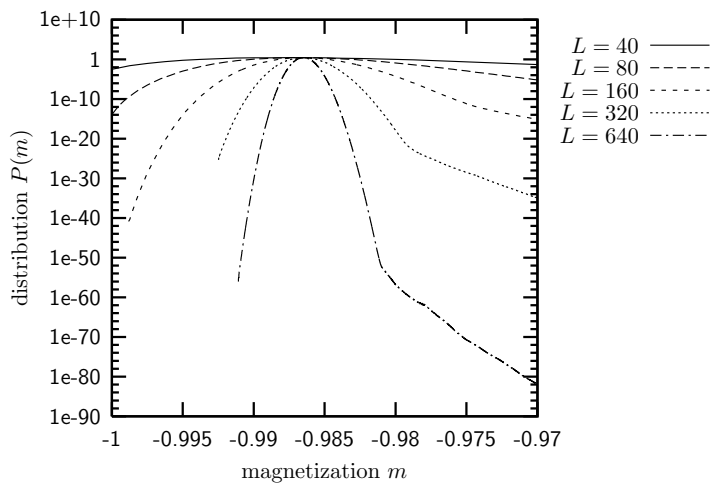}
  \end{center}
\vspace*{-0.2cm}
  \caption{Distribution of the magnetisation for the two-dimensional
    Ising model for different system sizes $L$ at
    the temperature $T=1.5$. The cusp indicates the transition region.
    On the left side of the cusp (evaporated system) a Gaussian peak
    is clearly visible, while on
    the right side of the cusp (condensed system) the stretched exponential
    behaviour 
    can be seen.}
  \vspace*{-1mm}
  \label{fig:cusp}
\end{figure}

\begin{figure}
  \begin{center}
    \includegraphics[clip,scale=0.85]{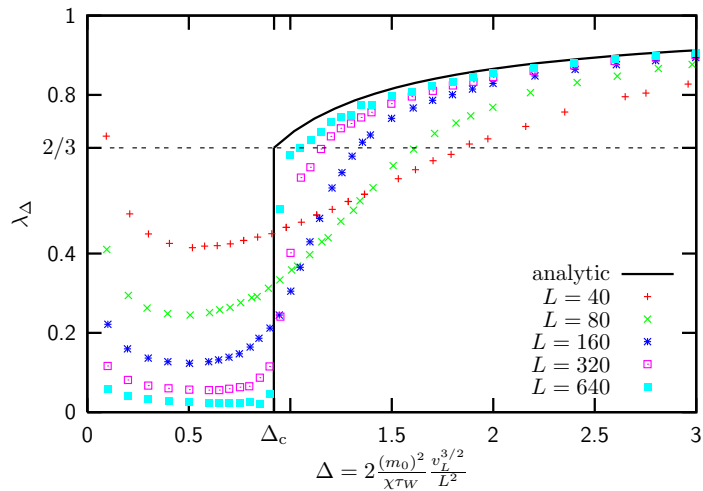}
  \end{center}
\vspace*{-0.4cm}
  \caption{Two-dimensional Ising model with nearest-neighbour interaction
    on a square lattice ($L=40, 80, \dots, 640$) at the temperature
    $T=1.5 \approx 0.7 \, T_c$. The error bars are not plotted since their 
    size is much smaller than that of the data symbols.}
  \label{fig:main.result}
\end{figure}

Figure~\ref{fig:cusp} shows the result of our ``Multimagnetic''
(Multicanonical for the magnetisation) simulations. The distribution of
the magnetisation shows at larger lattices sizes a cusp which divides
the evaporated and condensed region. Within the condensed region it
has a Gaussian form according to Eq.~(\ref{eqn:fluctuations}) while in
the condensed region a stretched exponential behaviour is visible, cf.
Eq.~(\ref{eqn:droplet}). All data points in Fig.~\ref{fig:main.result}
are in the vicinity of this cusp.
To get a feeling how the different configurations actual look like,
Figs.~\ref{fig:ce.a} and \ref{fig:ce.b} show an evaporated and
condensed system, respectively. Both systems have the same
magnetisation which was chosen to be the one at the transition point.
They represent extremes since during one simulation run these are the
configurations where the largest droplet is either minimal or maximal.

Figure~\ref{fig:main.result} shows the results for the fractions
$\lambda_\Delta$ for various lattice sizes. The solid line represents
the analytical value (result of the minimisation of
Eq.~(\ref{eqn:minimisation})).  Clearly, for larger lattice sizes the
result of the simulations approach the theoretical values nicely. The
jump from $\lambda_\Delta \approx 0$ to $\lambda_\Delta \approx 2/3$
at $\Delta_\mathrm{c} \approx 0.92$ confirms the theoretical
prediction that at the evaporation/condensation transition only $2/3$
of the excess of the magnetisation goes into the droplet.
%

The increase of $\lambda_\Delta$ for $\Delta \to 0$ can be explained
by the fact, that the minimal cluster size is $1$ (and not an
arbitrarily small fraction) but the excess that can be fixed is smaller
than $1$.

\section{Conclusion}
Our Monte Carlo data clearly confirm the theoretical results of Biskup
{\it et al.} \cite{kotecky}. The observed finite-size scaling behaviour 
fits perfectly with their predictions. All simulations were performed 
in thermal equilibrium and the abundance of droplets of intermediate 
size could be confirmed.
At the moment, additional simulations for different models are performed
that should prove the universal aspects of the theory. 

We would like to thank 
Roman Koteck\'y and
Thomas Neuhaus
for useful discussions.

\bibliographystyle{JHEP}
\bibliography{literature}

\end{document}